# Fast ungated five-dimensional cardiac MRI on a 1.5 T MR-linac for MRI-guided radiotherapy

M.L. Terpstra[1,†], T.E. Olausson[1,†], M.M.N. Aubert[2], C. Beijst[2], A. Sbrizzi[1], C.A.T. van den Berg[1], M.F. Fast[2]

[1]Computatinal Imaging Group for MRI Therapy & Diagnostics, Center of Image Sciences, University Medical Center Utrecht, Utrecht, the Netherlands

[2]Department of Radiotherapy, Imaging and Oncology Division, University Medical Center Utrecht, Utrecht, the Netherlands

† Equally contributing first authors

## Abstract

**Background:** Stereotactic arrhythmia radiotherapy (STAR) for patients with ventricular tachycardia is currently limited by complex cardiorespiratory motion. Hence, patient-specific five-dimensional magnetic resonance imaging (5D-MRI) motion models are required to overcome this clinical challenge. However, current 5D-MRI methods are hampered by long acquisition and reconstruction times, limiting their clinical availability for STAR treatments.

**Objective**: To develop and validate a fast, ungated 5D-MRI reconstruction method, providing personalized motion characterization to support MRI-guided STAR treatments.

**Methods:** We propose a fast, ungated 5D-MRI reconstruction method based on the previously established cardiac MR-MOTUS framework. The method utilizes a variable-density 3D Cartesian acquisition with a joint optimization framework to reconstruct a motion-corrected reference image and low-rank deformation vector fields (DVFs). By exploiting the low rank structure of the DVFs, we explicitly disentangle respiratory and cardiac motion during optimization. After optimization, the DVFs are used as navigators for cardiorespiratory motion, enabling fast 5D-MRI reconstruction with a retrospectively adjustable number of motion states. Validation was performed using digital and physical cardiorespiratory phantoms. Furthermore, the approach was evaluated in vivo using 10 healthy volunteers, comparing motion consistency with standard 2D cine MRI.

**Results:** Validation of 5D CMR-MOTUS using digital and physical phantoms demonstrated accurate 5D-MRI reconstruction. In the physical phantom, 5D CMR-MOTUS achieved a left-ventricle DICE coefficient of 0.96 ± 0.01. In the volunteer cohort, the 5D-MRI scans showed strong motion consistency when compared to 2D cine MRI, yielding a cardiac motion amplitude error of 0.1 ± 0.9 mm and a respiratory motion amplitude error of 0.2 ± 2.9 mm. Crucially, 5D-MRI data were acquired in one minute and reconstructed in six minutes.

**Conclusions:** The proposed 5D-MRI method enables rapid, high-quality, and personalized motion characterization, demonstrating potential for integration into MRI-guided STAR treatments.

**Data Availability:** The 3D k-space data, 5D reconstructions, 2D cine imaging, and manual segmentations for the ten volunteers are publicly available at https://doi.org/10.5281/zenodo.21278894.

# Introduction

Stereotactic arrhythmia radiotherapy (STAR) is an emerging, non-invasive last-resort treatment modality for patients with ventricular tachycardia (VT). By delivering a single high-fraction dose of radiation to the myocardium, re-entrant electric signals that induce VT are effectively terminated [1–3]. However, the treatment effectiveness may be limited by substantial cardiorespiratory motion, the small target size, and the radiosensitive nature of cardiac tissue [4–6]. To maximize treatment efficiency and minimize radiation toxicity, accurate motion management is paramount when performing STAR treatments. Hybrid MRI-linac systems address these challenges by leveraging the excellent soft-tissue contrast of magnetic resonance imaging (MRI) for online treatment planning and real-time motion management [7–9]. Recently, the feasibility of using MRI-guided STAR treatments using MR-linac devices has been demonstrated [10,11]. However, one major challenge is cardiorespiratory motion. Current clinical practice of MRI-guided radiotherapy uses respiratory gating based on 2D cine imaging, limiting motion accuracy and treatment efficiency. Recently, it has been demonstrated that 4D respiratory-resolved MRI enables highly accurate volumetric tracking of mobile structures [12]. However, 4D-MRI cannot resolve the cardiac motion, which must be accounted for during STAR treatments. While 3D cine imaging could be leveraged to reconstruct all physiological motion [13], cardiac and respiratory motion must be individually characterized to accurately estimate treatment margins[14], and to enable flexible treatment regimens (e.g., respiratory tracking and cardiac gating) to mitigate target motion [15]. Hence, a patient-specific motion model based on accurate five-dimensional cardiothoracic-resolved MRI (5D-MRI) must be available for treatment planning. Moreover, 5D-MRI allows for mitigation of cardiac and respiratory motion during STAR treatment by gating or tracking, enabling sparing of radiosensitive healthy tissue[16].

As the treatment starts immediately after planning in an adaptive MR-linac setting, accurate 5D-MRI must be acquired and reconstructed within a 10-minute pre-treatment window, spanning from the start of imaging after patient positioning to the availability of the 5D-MRI in the treatment planning system, to maximize target accuracy, maintain patient comfort, and ensure high treatment efficiency [17,18]. However, obtaining 5D-MRI within this time window is challenging. Recent 5D-MRI methods propose utilizing a free-running, i.e., free-breathing and without ECG, non-Cartesian acquisition that leverages a self-navigation signal to sort the acquired k-space into distinct motion states [19–21]. By sorting all k-space samples into cardiothoracic motion states according to these self-navigation signals, images are reconstructed using a custom compressed sensing (CS) algorithm.

While these methods can produce high-quality 5D-MRI, the slow acquisition speed of MRI leaves the cardiothoracic motion states severely undersampled, resulting in a highly ill-posed reconstruction problem. Typically, an acquisition time of approximately 5-15 minutes is used to ensure sufficient coverage of every cardiorespiratory motion state [19–22]. Furthermore, non-Cartesian acquisitions require repeated gridding steps during reconstruction, which in turn require expensive interpolation operations [23]. This results in long reconstruction times up to several hours. Due to these limitations, 5D-MRI based on non-Cartesian acquisitions and compressed sensing is currently not suitable for MRI-guided STAR treatments.

Previous studies demonstrated that explicitly modeling motion using deformation vector fields (DVFs) is significantly more data-efficient and compressible than directly modeling motion-resolved images [21,24], and has been applied to 3D time-resolved positron emission tomography (PET)[25] and cone-beam CT (CBCT) imaging[26,27]The CMR-MOTUS framework jointly reconstructs high-quality 3D time-resolved DVFs and a motion-corrected reference image using a Cartesian acquisition[13]. The method is highly data-efficient: by correlating all motion events over time-resolved DVFs to reconstruct a single motion-corrected reference image, it effectively uses all data points and improves the SNR of the reference image. Additionally, it exploits the inherent compressibility of DVFs by reconstructing an explicit low-rank factorization in space and time [28], efficiently capturing the underlying spatiotemporal correlations of cardiorespiratory motion.

In this work, we propose using CMR-MOTUS with a Cartesian acquisition to obtain 5D-MRI for STAR treatment planning within 10 minutes on an MR-linac. To achieve this, we leverage a low-rank motion model and explicitly disentangle the cardiac from the respiratory motion. This structure provides data-driven navigators to construct the cardiorespiratory motion states. We investigate the performance of 5D CMR-MOTUS on digital and physical phantoms to represent cardiorespiratory motion as a function of the k-space sampling speed, image resolution, and scan time. Finally, we evaluate the performance of 5D CMR-MOTUS in ten healthy volunteers.

# Methods

Here, we describe the basis of the CMR-MOTUS algorithm, the extraction of cardiorespiratory surrogate signals from motion fields, and the construction of the 5D motion states.

## Reconstruction Framework

In CMR-MOTUS, we jointly solve for a motion-corrected reference image and time − resolved deformation vector fields (DVFs) by solving the following minimization problem:

$$[\boldsymbol{q}^*, \boldsymbol{D}^*] = \operatorname{argmin}_{\boldsymbol{q},\boldsymbol{D}} \sum_{t=1}^{T} \sum_{c=1}^{N_c} ||\mathcal{M}_t \boldsymbol{s}_{t,c}(\mathbf{q}, \mathbf{D}_\mathrm{t}) - \boldsymbol{m}_{t,c}||_2^2 + \mathcal{R}(\boldsymbol{q}, \boldsymbol{D}). \quad (1)$$

Here, $\boldsymbol{m}_{t,c}$ is the measured k-space for receive coil $c \in \{1, \dots, N_c\}$ at timepoint $t$, $\boldsymbol{q}$ is the motion-corrected reference image, $\boldsymbol{D} \in \mathbb{R}^{T \times 3N^3}$ are the 3D time-resolved DVFs, $\mathcal{M}_\mathrm{t}$ is the sampling mask containing the k-space locations that were sampled in $\boldsymbol{m}_{t,c}$, and $\mathcal{R}$ is any regularization that could be applied to $\boldsymbol{q}$ or $\boldsymbol{D}$ to reduce the ill-posedness of the minimization problem, respectively. $\boldsymbol{s}_{t,c}(\boldsymbol{q}, \boldsymbol{D}_t) = FC_c(W(\boldsymbol{q}, \boldsymbol{D}_t))$ is signal estimated at timepoint t for coil c by applying the CMR-MOTUS forward operation, which includes the Fourier transform operator F, multiplication with the coil sensitivity maps operator $C_c$, and image warping operator $W$, which warps $\boldsymbol{q}$ with deformation vector field $\boldsymbol{D}_t$ [13]. This optimization procedure is illustrated in Figure 1.

### Motion field representation

Due to the large number of unknown parameters in equation 1, this joint optimization cannot be performed directly. Huttinga et al. have shown that the number of unknowns of $\boldsymbol{D}$ can be significantly reduced by modeling the DVFs using cubic B-splines, exploiting the spatial smoothness of physiological motion. Second, we employ a low-rank motion basis [28], further taking advantage of the spatiotemporal present in cardiorespiratory motion to reduce the complexity of the inverse problem. Hence, instead of representing the DVF as $\boldsymbol{D} \in \mathbb{R}^{T \times 3N^3}$, where $T$ is the number of timesteps and $N^3$ is the number of voxels in 3D, we approximate the full DVFs as $\boldsymbol{D} \approx \boldsymbol{\Phi\Psi}^T$ with spatial basis $\boldsymbol{\Phi} \in \mathbb{R}^{3LR}$ and temporal basis $\boldsymbol{\Psi} \in \mathbb{R}^{TR}$ such that $3LR + TR \ll 3TN^3$. Here, $R$ is the explicitly chosen rank of the low-rank basis, and $L$ is the number of cubic B-splines used in the spatial basis representation [24].

### Estimating motion states

Typically, $\boldsymbol{D}$ contains the mixed cardiorespiratory dynamics. However, to obtain 5D-MRI, it is crucial to separate the respiratory from the cardiac motion dynamics. Therefore, it is natural to take advantage of the low-rank structure of $\boldsymbol{\Phi\Psi}^T$ to ensure the cardiac and respiratory motion are separated in the deformation vector fields, treating $\boldsymbol{\Psi}$ as navigator signals for physiological motion. Then, $\mathbf{\Psi}$ and $\boldsymbol{q}$ are used to retrospectively construct the cardiorespiratory motion states, yielding 5D-MRI.

One way to achieve this disentanglement is to recognize that the frequency ranges of respiratory motion (generally 0.2-0.4 Hz) and cardiac motion (generally 0.8-2 Hz) do not overlap [19]. Therefore, we can partition $\mathbf{\Psi}$ into $\mathbf{\Psi} = \begin{bmatrix} \mathbf{\Psi}_{\text{resp}} \\ \mathbf{\Psi}_{\text{cardiac}} \end{bmatrix}$ such that $N$ components of the low-rank DVF approximation represent low-frequency motion typically associated with respiratory motion and $M$ components representing high-frequency motion typically associated with cardiac motion ($N + M \leq R$). Consequently, $\mathbf{\Phi}$ is decomposed into $\mathbf{\Phi}_{\text{resp}}$ and $\mathbf{\Phi}_{\text{cardiac}}$, respectively. This way, an explicitly disentangled motion basis is created for respiratory and cardiac motion.

## Data sorting

Given that we have $\boldsymbol{q}$, $\mathbf{\Phi}_{\text{resp}}$, $\mathbf{\Phi}_{\text{cardiac}}$, $\mathbf{\Psi}_{\text{resp}}$, and $\mathbf{\Psi}_{\text{cardiac}}$, we can reconstruct the 5D cardiothoracic-resolved MRI (Figure 1). To reconstruct the 5D motion states, we first perform binning on $\mathbf{\Psi}_{\text{resp}}$, and $\mathbf{\Psi}_{\text{cardiac}}$, sorting the temporal navigators into $\mathbf{B}_{\text{resp}}$ and $\mathbf{B}_{\text{cardiac}}$ bins. The number of bins can be chosen retrospectively after $\mathbf{\Psi}$ is reconstructed. In this work, we performed hybrid binning to account for the hysteresis of respiratory and cardiac motion [29]. Using these binned temporal signals, we construct the 5D DVFs by multiplying $\mathbf{\Phi}_{\text{resp}}$ and $\mathbf{\Phi}_{\text{cardiac}}$ with the respective binned navigator values. The full 5D-MRI is constructed by warping $\boldsymbol{q}$ with each 5D DVF for every motion state. This is a major difference compared to methods that sort k-space into cardiorespiratory motion states prior to reconstruction, as our proposed method enables retrospective selection of $\mathbf{B}_{\text{resp}}$ and $\mathbf{B}_{\text{cardiac}}$ with constant image quality.

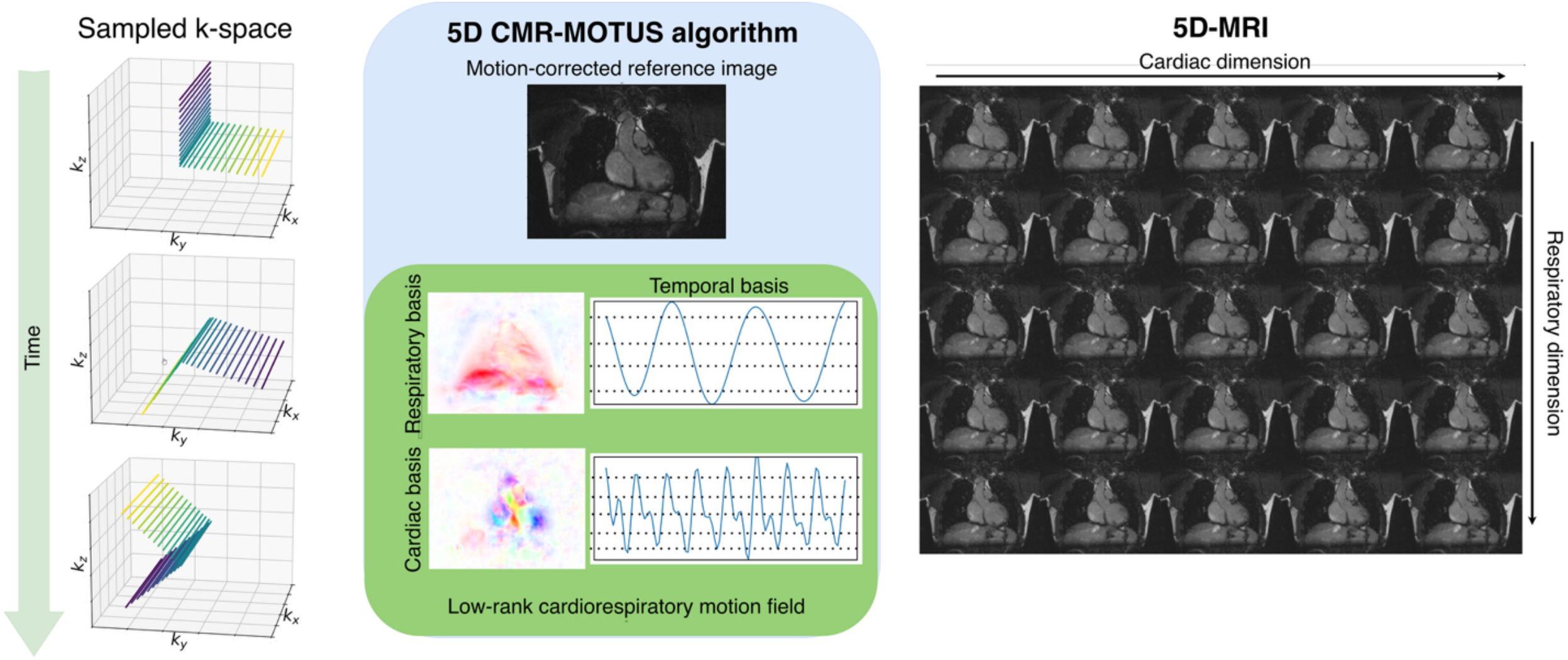

***Figure 1: Method overview**. To reconstruct our 5D-MRI, free-running time-resolved k-space is acquired using the OPRA sampling pattern (left). Using the 5D CMR-MOTUS algorithm (middle), the motion-corrected reference image and a low-rank approximation of time-resolved DVFs (green), consisting of explicitly disentangled spatial (**$\mathbf{\Phi}$**) and temporal*

*($\boldsymbol{\Psi}$) bases, are jointly optimized to minimize the difference to the sampled k-space. After optimization, $\boldsymbol{\Psi}$ is used as a cardiorespiratory navigator, and binned into discrete motion states. Then, 5D-MRI (right) is constructed by warping the motion-corrected reference image with the binned low-rank DVF. The number of bins can be chosen retrospectively. Here, 5 respiratory and 5 cardiac phases are shown.*

## Acquisition & implementation

In this study, we leverage a Cartesian sampling pattern. While traditionally non-Cartesian sampling patterns are used for dynamic imaging, we employ the Ordered Pseudo-RAdial (OPRA) sampling trajectory, a Cartesian sampling strategy optimized for temporal imaging using balanced steady-state free precession (bSSFP) sequences [30]. The OPRA sampling pattern arranges the phase encodes per shot in a leaflet (Figure 1). The number of phase encodes per leaflet is a user-defined parameter modulating the temporal resolution per shot. In this study, each OPRA leaflet comprises 26 phase-encode readouts (TR = 3.8 ms). As each motion state corresponds to one leaflet, the MRI signal is acquired with a sampling rate of approximately 10 Hz, providing a balance between temporal resolution and sampling density per leaflet. The specific sampling parameters per experiment are described in Table 1.

The 5D CMR-MOTUS framework was implemented by extending the CMR-MOTUS framework[13]. Using the autodifferentiation capabilities of the PyTorch framework, we perform joint stochastic optimization in time of the reference image $\boldsymbol{q}$ and the low-rank deformation motion field components $\boldsymbol{\Phi}\boldsymbol{\Psi}^T$ for 15 epochs using minibatches of 20 leaflets [30]. The reference image was initialized using the zero-filled inverse Fourier transform of the k-space data, and coil sensitivity maps were estimated using ESPIRiT [31]. The spatial component basis $\boldsymbol{\Phi}$ is optimized using a coarse-to-fine approach: the spatial basis initially uses 15 splines in each spatial dimension ($L = 15 \times 3$), linearly increasing to d/2 splines over 10 epochs, where d is the number of voxels in that dimension. The spatial spline basis is computed using an efficient spline interpolation library [32]. For the temporal basis, no spline basis was used. In total, our low-rank motion model consists of 9 components that represent physiological motion by temporal bandpass filtering: three components are associated with $\boldsymbol{\Phi}_{\text{resp}}$, containing temporal frequencies between 0.1 and 0.4 Hz, five components are associated with $\boldsymbol{\Phi}_{\text{cardiac}}$, containing temporal frequencies between 0.7 and 2 Hz, and one component accounts for drifting motion and contains frequencies below 0.1 Hz. This number of components were demonstrated to successfully capture respiratory and cardiac motion [24]. To regularize the reconstruction problem and improve the image quality, we apply an L1 wavelet penalty on the reference image ($\lambda = 1e^{-1}$). With these regularizations applied, equation 1 is minimized using the Adam optimizer [33] with a learning rate of 0.02 for $\boldsymbol{\Phi}$ and $\boldsymbol{\Psi}$, and 0.075 for the reference image. An exponential learning rate schedule with γ = 0.8 was applied. Optimization hyperparameters and spline initialization settings were tuned using automated Bayesian optimization (Optuna) on a single 60-second volunteer acquisition by minimizing the unsupervised k-space data-consistency residual. A hyperparameter analysis on the number of motion field components and regularization parameters is presented in Supplementary Information Document 1. Once established, all hyperparameters were strictly fixed across all evaluations of digital phantoms, physical phantoms, and volunteers. All experiments were performed using a NVIDIA L40S GPU with 48GB VRAM.

## Experiments

We performed three experiments to validate the 5D CMR-MOTUS performance: a digital phantom experiment, a physical phantom experiment, and an in-vivo 5D reconstruction.

### Digital phantom

To quantitatively evaluate the performance of the proposed method, cardiorespiratory motion was simulated using the XCAT numerical phantom [34]. Using XCAT, a torso was simulated at 1mm$^3$ isotropic resolution, with cardiac motion at a frequency of 1 Hz and respiratory motion with a 0.2 Hz frequency. The respiratory motion amplitude was 10 mm in the anterior-posterior direction and 25 mm in the feet-head direction. The MRXCAT framework was used to simulate the contrast of a bSSFP acquisition (TR=3.8 ms, FA=50°) [35].

To avoid simulating MRXCAT for every TR, respiratory and cardiac deformation vector fields were simulated separately by XCAT to serve as spatial bases. These spatial bases were used with a custom waveform describing the temporal motion pattern at high temporal resolution, enabling sparse motion simulation at TR-resolution. Cardiac motion was modulated using a sinusoid with a frequency of 1 Hz, while respiratory motion was modulated using a $\cos^4$ function of 0.2 Hz. To investigate the performance of our proposed method, we simulated MRXCAT at 2, 3, and 4 mm$^3$ resolution for 30, 60, 90, and 120 seconds. The temporal resolution of these simulations ranged from 3 to 10 Hz.

A Cartesian dynamic acquisition was simulated using the OPRA sampling pattern [30]. Using a TR of 3.8 ms, this resulted in 26 phase encodes on a single leaflet at 10 Hz, and 88 phase encodes on a leaflet at 3 Hz.

The reconstructions generated by 5D CMR-MOTUS were evaluated by comparing image quality, as measured by PSNR, with the ground-truth 5D reconstruction based on the MRXCAT deformation vector fields. Moreover, we evaluated the end-point error (EPE) $EPE = ||\mathrm{DVF}_{true} - \mathrm{DVF}_{estimated}||_2$ between the estimated 5D DVFs and ground-truth MRXCAT 5D DVFs.

### Physical phantom reconstruction

A deformable cardiac motion phantom [36] was attached to a motor driving a piston. Then, both were placed on a rigid motion platform. This setup was scanned on the 1.5 T Unity MR-linac system (Elekta AB, Stockholm, Sweden) using an 8-channel phased-array coil (Figure 2). The motion phantom was driven with sinusoidal waveforms, using a 1Hz sinusoidal waveform with a peak amplitude of 10 mm for cardiac motion, and a 0.4 Hz sinusoidal waveform with a 20 mm peak amplitude for the respiratory motion. The dynamic scan used a 3D bSSFP sequence with TR/TE = 3.7/1.9 ms, flip angle = 50°, matrix size 384 × 128 × 149, and 49998 Cartesian phase-encode readouts (acquisition time approximately 3 minutes) to investigate scan-time performance. In addition to the time-resolved dynamic acquisition, four static 3D spoiled gradient-echo scans were acquired as ground truth at extreme motion setpoints. These data were acquired with a TR/TE = 3.449/1.613 ms, flip angle = 10°, matrix size 384 × 128 × 149, and 49998 readouts (scan duration roughly 172 seconds). For all five scans, the acquisition voxel size was isotropic 2.0 mm. Figure 2 shows the experiment setup.

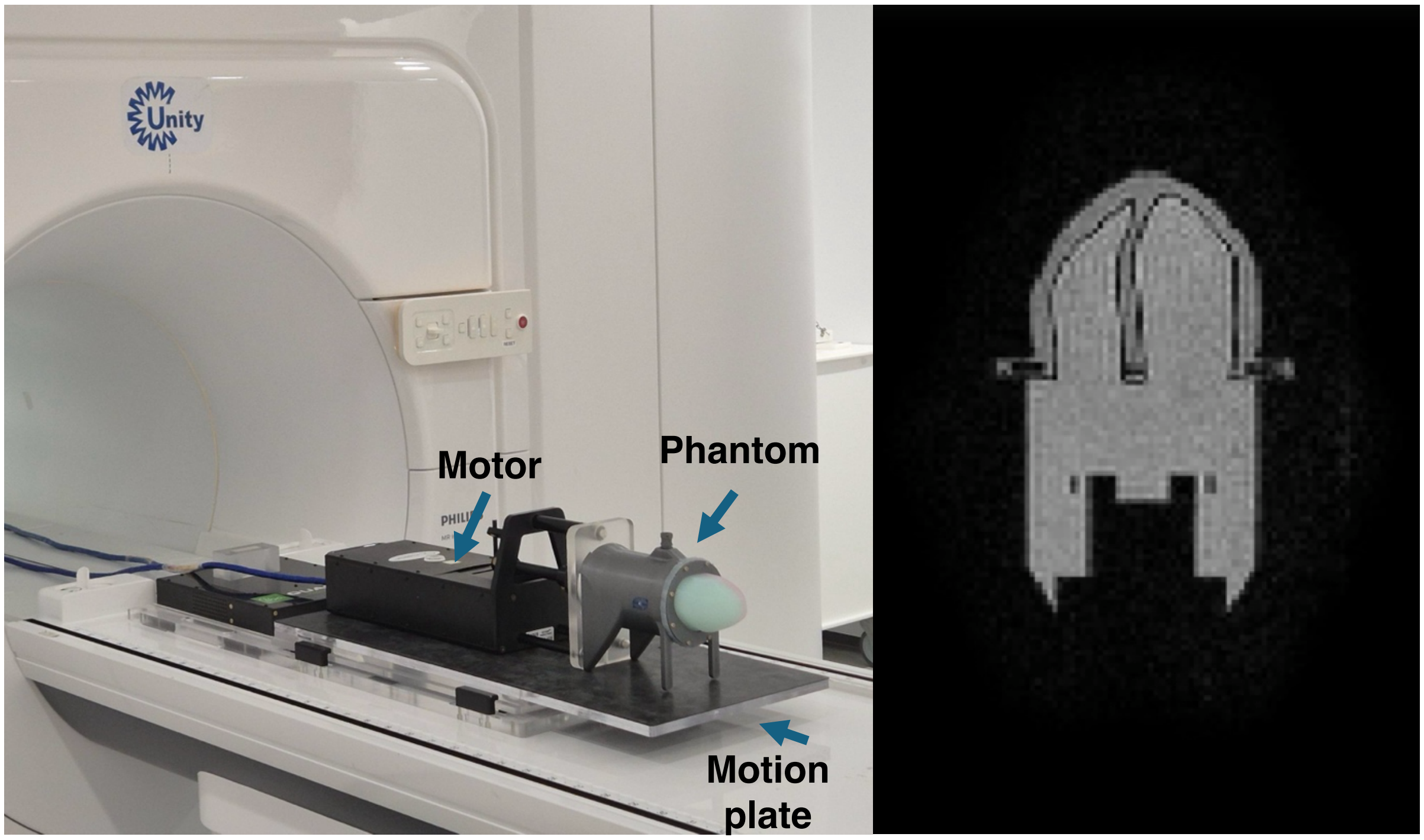

**Figure 2: Experiment setup.** The deformable cardiac phantom is placed onto a motion platform, enabling simultaneous cardiac and respiratory motion (left). The coil was placed over the phantom "ventricles", while the motor drives the phantom with 0.4Hz 20 mm respiratory amplitude, and 1Hz 10 mm cardiac deformation amplitude. An example static coronal MRI scan of the phantom is shown on the right.

The dynamic data were reconstructed using two separate algorithms. First, a baseline compressed-sensing reconstruction [20] was created by performing retrospective 5D traditional binning using the ground-truth waveforms. In total, 10 cardiac bins and 4 respiratory bins were used to reconstruct these data. We employed an XD-GRASP reconstruction [20] with temporal total variation along the respiratory and cardiac dimensions ($\lambda_{resp} = \lambda_{cardiac} = 0.01$) using the BART toolbox [37] (Version 1.0.00). These regularization parameters were found by performing an independent grid search over the respiratory and cardiac regularization strength parameters ($\lambda \in [0.003, 0.01, 0.03]$, evaluating the reconstruction quality using the L2 norm on held-out k-space.

Second, a 5D CMR-MOTUS reconstruction was created using the first 15912 phase encodes (approximately one minute of data), and a 5D CMR-MOTUS reconstruction was made using all 49998 phase encodes. A rank-three cubic B-spline basis was used to model the respiratory deformation, while a rank-five cubic B-spline basis was used to model the cardiac deformation.

To evaluate the reconstruction quality, we manually segmented the left and right ventricles using nnInteractive [38] in the static reference images, in the extreme motion phases of the XD-GRASP reconstruction, and in the 5D CMR-MOTUS reference image. The CMR-MOTUS segmentations were then propagated to the extremal positions using the 5D motion fields. These contours were compared with the static reference image contours using the Dice similarity metric.

## In-vivo MRI reconstruction

In-vivo MRI was obtained in 10 volunteers on a 1.5 T Unity MR-linac using an eight-channel phased-array coil. Informed consent was obtained from each subject in accordance with institutional guidelines and Local Ethics Committee approval (NL59820.041.7). First, a coronal 2D cine MRI was obtained to validate the liver dome position and the cardiac dynamic amplitude. The relevant scan parameters for these 2D cine MRI are listed in Table 1. To obtain 2D cardiorespiratory deformation vector fields, groupwise registration was performed on the 2D cine images to obtain time-resolved 2D deformation vector fields [39]. Separate respiratory and cardiac DVFs were then obtained by performing singular value decomposition on these time-resolved DVFs, retaining the first and second principal components, respectively. The reference image of this groupwise registration was manually delineated using nnInteractive [38], segmenting the left ventricle, right ventricle, myocardium, liver, left lung, and right lung.

Next, three-dimensional free-breathing MRI was acquired without external navigators or triggering. The data were acquired using the OPRA sampling pattern, with the relevant scan parameters listed in Table 1. While both bSSFP and exploratory non-contrast spoiled GRE acquisitions were collected, the 5D CMR-MOTUS reconstruction pipeline in this study was developed and validated exclusively on the bSSFP acquisitions. The field of view was chosen to include the body contour in transverse view, which is essential for dose calculation when considering STAR treatments. In total, approximately 15800 k-space phase encodes were obtained for each volunteer for each 5D-MRI acquisition and reconstructed using the 5D CMR-MOTUS algorithm into 10 respiratory motion states and 10 cardiac motion states. Moreover, to demonstrate the ability of retrospective selection of the number of phases, we also reconstructed the same data into 20 respiratory motion states and 20 cardiac motion states. These k-space and reconstructions are publicly available at https://doi.org/10.5281/zenodo.21278894. The reference image of these reconstructions was manually delineated using nnInteractive, segmenting the left ventricle, right ventricle, myocardium, liver, aorta, left lung, and right lung.

To evaluate the motion quality of the 5D-MRI, we computed the respiratory and cardiac amplitude within the same coronal slice as the 2D cine acquisitions. To this end, we compute $\mathrm{DVF}_{\mathrm{Max,resp.}} = \mathrm{DVF}_{\mathrm{Exhale}} - \mathrm{DVF}_{\mathrm{Inhale}}$ and $\mathrm{DVF}_{\mathrm{Max,cardiac}} = \mathrm{DVF}_{\mathrm{Systole}} - \mathrm{DVF}_{\mathrm{Diastole}}$ . The respiratory amplitude was then defined as the 95$^{\mathrm{th}}$ percentile of $\mathrm{DVF}_{\mathrm{Max,resp.}}$ within the liver contour of the 2D and 5D acquisition. Similarly, the cardiac amplitude was defined as the 95$^{\mathrm{th}}$ percentile of $\mathrm{DVF}_{\mathrm{Max,cardiac}}$ within the myocardium contour. Moreover, we analyzed the bias and standard deviation of the mean difference between the 2D and 5D reconstructions using Bland-Altman analysis. Finally, we quantified the breath-to-breath variability within the 5D acquisition and tested for changes in breathing between the 2D cine and the 5D acquisition. For the variability, we determined the end-inhale displacement of every complete breath in the 5D acquisition and report its mean ± standard deviation and coefficient of variation (CV). For the comparison, we considered all overlapping segments of the 5D time series with the same duration as each volunteer's 2D cine, one starting at each time point (98.8-ms spacing), and computed the displacement within each segment as for the 2D cine (deepest inhale relative to deepest exhale). The 2D cine displacement was expressed as its percentile rank among these duration-matched segments and as its excess (mm) over their median.

***Table 1: Sequence parameters****. Relevant scan parameters for three experiments. The "2D-cine MRI" column depicts the sequence parameters used to acquire the validation 2D cardiac cine in coronal orientation. Then, two OPRA acquisitions are listed for the 5D-MRI with bSSFP and GRE contrast, respectively. 5D CMR-MOTUS was only validated on the bSSFP contrast. All acquisitions were performed on a 1.5 T MR-linac.*

| | | 5D-MRI | |
|---|---|---|---|
| Parameter | 2D Cine MRI | OPRA bSSFP | OPRA GRE |
| Pulse Sequence | Balanced gradient echo | Balanced steady-state free precession | Spoiled gradient-echo |
| Readout direction | FH | RL | RL |
| Number of coils | 8 | 8 | 8 |
| Slice thickness (mm) | 9 | - | - |
| TR/TE (ms) | 3.2/1.6 | 3.8/1.9 | 3.4/1.6 |
| Readouts per frame | 38 | 26 | 26 |
| Temporal resolution (Hz) | 8.3 | 10.1 | 11.3 |
| Flip angle (°) | 50 | 50 | 10 |
| Bandwidth (Hz/px) | 1102 | 1293 | 1293 |
| FOV (mm) | 288 × 256 | 248 × 200 × 385 | 248 × 200 × 385 |
| Spatial resolution (mm) | 1.6 ×1.6 | 2 ×2 ×2 | 2 ×2 ×2 |
| Matrix size | 180 ×160 | 124 ×100 ×193 | 124 ×100 ×193 |
| Scan time (s) | 10 | 180 (60 s used for 5D CMR-MOTUS) | 180 |

# Results

### Digital phantom

An example reconstruction from the 5D CMR-MOTUS reconstruction, compared with the ground truth, is shown in Figure 3, demonstrating the method's accurate motion correction and motion estimation capabilities, as indicated by the sharp structures and the correspondence of anatomical structure locations. The simulated k-space was reconstructed in approximately eight minutes on a single NVIDIA L40S. When performing quantitative analysis of image quality relative to the ground truth (Figure 4), image quality increases with longer acquisition times, but shows diminishing returns beyond 60 seconds (PSNR 24 dB for 60 seconds versus 24.5 dB for 120 seconds). Regarding temporal resolution, the digital phantom indicates that 4Hz is sufficient to reconstruct an accurate 5D-MRI (Figure 3). The image quality is substantially lower at 4 $mm^3$ compared to higher resolutions (PSNR 22.8 dB versus >= 24.5 dB). At a spatial resolution of 2 $mm^3$ and 60-second acquisition duration (10 Hz), 5D CMR-MOTUS achieved a sub-millimeter EPE of $0.99 \pm 0.03$ mm.

### Physical phantom

For the physical phantom, 5D CMR-MOTUS generates high-quality cardiorespiratory MRI with a one-minute acquisition, whereas the XD-GRASP reconstruction suffers from significant reconstruction artifacts with a three-minute acquisition. Moreover, the 5D CMR-MOTUS reconstructed phantom shows clear motion correction, indicated by the sharp piston attachment and sharp border between the fluid pool and outer shell. Moreover, there are no noticeable undersampling artifacts. The 5D CMR-MOTUS was reconstructed in 6 minutes, while the XD-GRASP took approximately 30 minutes. When comparing the left-ventricle DICE with the ground-truth, XD-GRASP achieves a DICE of 0.62 ± 0.05 and 5D CMR-MOTUS 0.96 ± 0.01, indicating accurate cardiorespiratory DVF reconstruction. For the right-ventricle DICE, XD-GRASP achieves a DICE of 0.73 ± 0.05 and 5D CMR-MOTUS 0.91 ± 0.01.

### In-vivo results

One minute of k-space data was reconstructed into 5D-MRI in 6 minutes. The short-axis orientations are shown in Figure 6.  Figure 7 shows the four-chamber, short-axis, and two-chamber views for one volunteer for the four extreme motion states, respectively. Despite the one-minute acquisition, minimal undersampling or motion artifacts are visible, indicating the favorable properties of the 5D CMR-MOTUS reconstruction framework. Moreover, a clear disentanglement is visible between the cardiorespiratory motion states.

Compared with the 2D cine acquisition, there is good correspondence between cardiac and respiratory motion in all volunteers, with a mean cardiac motion difference of -0.07 ± 1.7 mm, and a mean respiratory motion difference of 0.16 ± 5.5 mm (Figure 8). Bland-Altman analysis revealed that the cardiac motion amplitude error was 0.07 ± 0.9 mm, and the respiratory motion amplitude error was 0.16 ± 2.9 mm. The mean absolute error was 2.31 ± 1.47 mm for the respiratory motion and 0.8 ± 0.27 mm for the cardiac motion. Breath-by-breath analysis of the continuous 60-s 5D trajectories (Supplementary Table S1) showed low intra-scan respiratory variability, with end-inhale displacement varying with a mean CV of 8.1% and breathing period with a CV of 8.5%. However, in the duration-matched sliding-window analysis, the 2D cine respiratory displacement exceeded the 95th percentile of same-length 5D windows in 7 of 10 volunteers, exceeding the median 5D window by 2.9 ± 2.4 mm on average, indicating deeper breathing during the initial 2D cine acquisition than during the subsequent 5D scan.

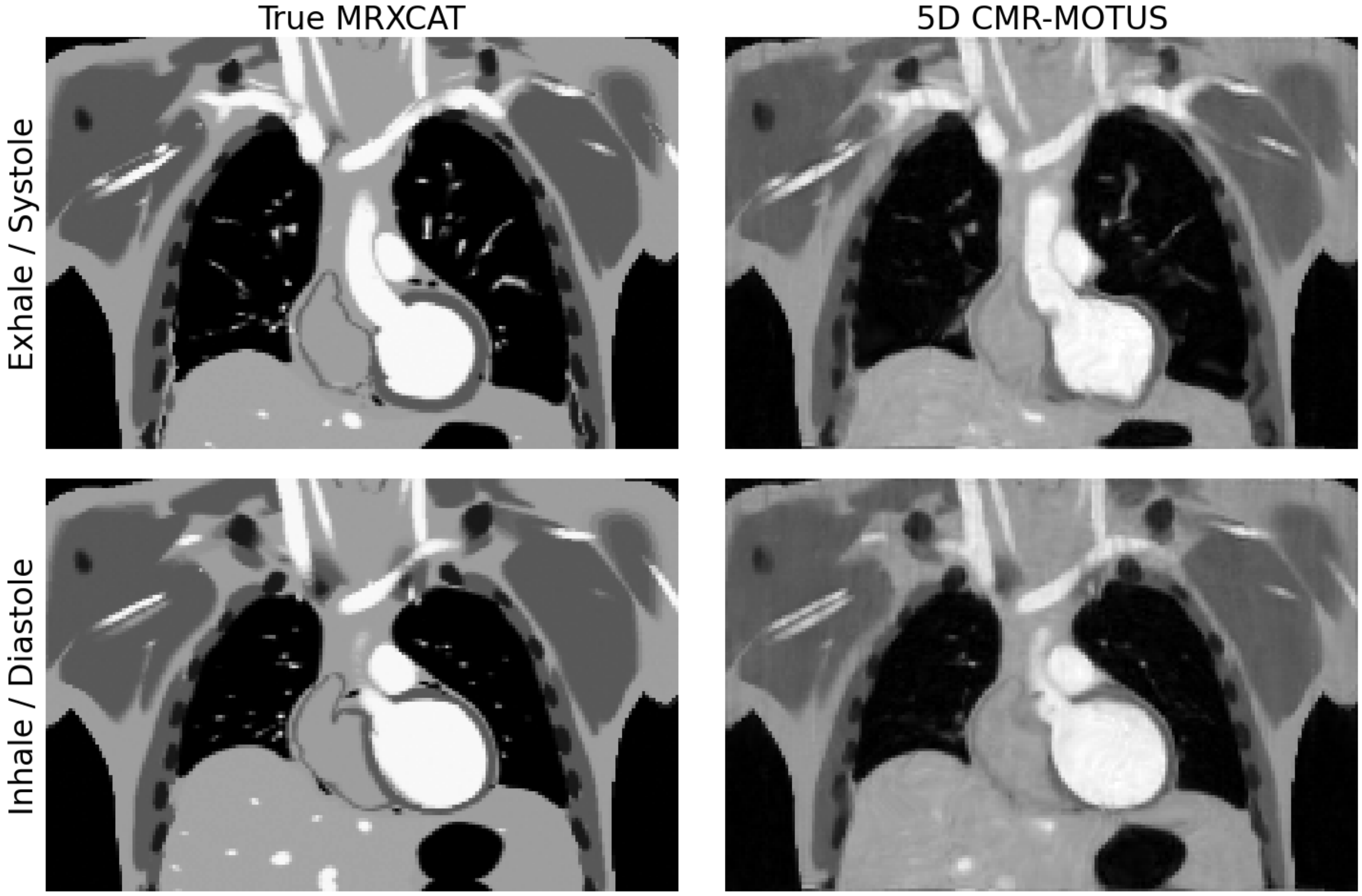

**Figure 3: Example MRXCAT reconstruction**. On the left, the ground-truth MRXCAT digital phantom simulations, while the right side shows 5D CMR-MOTUS reconstructions using a one-minute acquisition with a 10Hz temporal resolution.

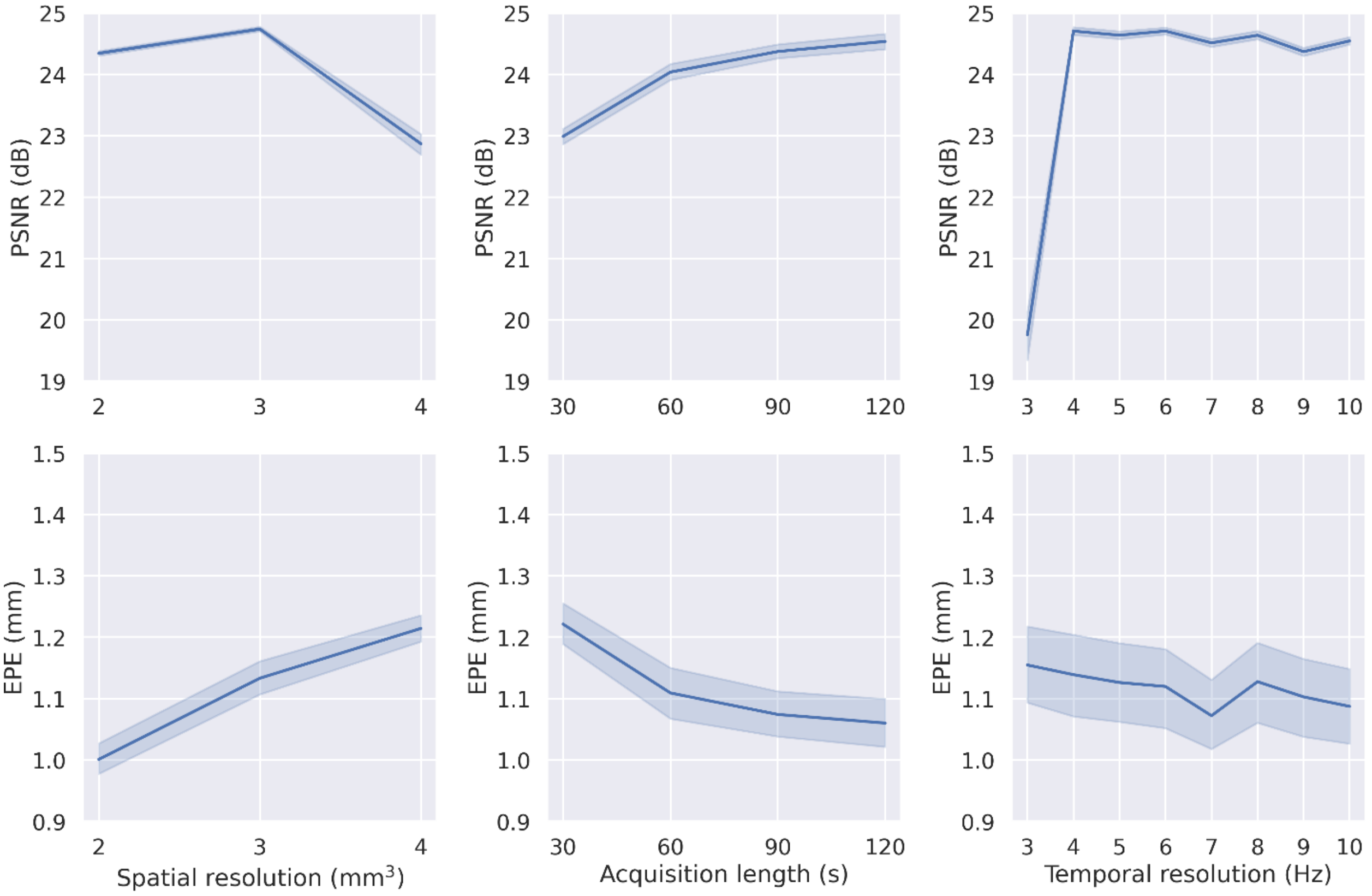

***Figure 4: Quantitative digital phantom results***. *Comparing the estimated 5D-MRI to the ground-truth MRXCAT 5D-MRI, the PSNR is minimal when the spatial resolution is* $> 3$ *mm*$^3$ *(top, left). The image quality significantly decreases when the acquisition is shorter than 60 seconds (top, middle). To resolve cardiac and respiratory motion, sufficient performance is obtained when the temporal resolution is at least 4 Hz (top, right). The EPE is most strongly affected by spatial resolution (bottom, left).*

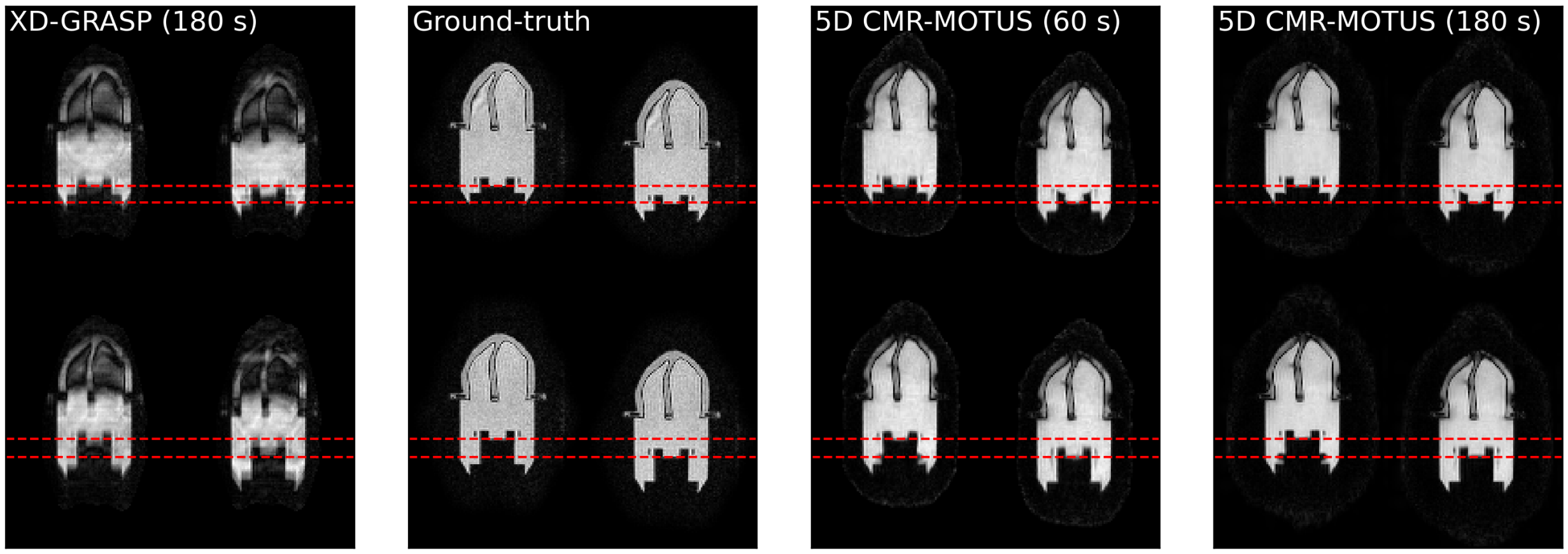

***Figure 5: Physical phantom results***. *Top row shows the images in systole, while bottom row shows diastole motion states. The columns of each reconstruction show exhale and inhale, respectively. The XD-GRASP reconstruction suffers from significant artifacts, whereas the 5D CMR-MOTUS accurately captures cardiorespiratory motion. Similar image quality is obtained for the one-minute 5D CMR-MOTUS acquisition and three-minute 5D CMR-MOTUS acquisition. The 5D CMR-MOTUS and XD-GRASP reconstructions suffer from banding artifacts due to the bSSFP acquisition compared to the GRE acquisition of the reference.*

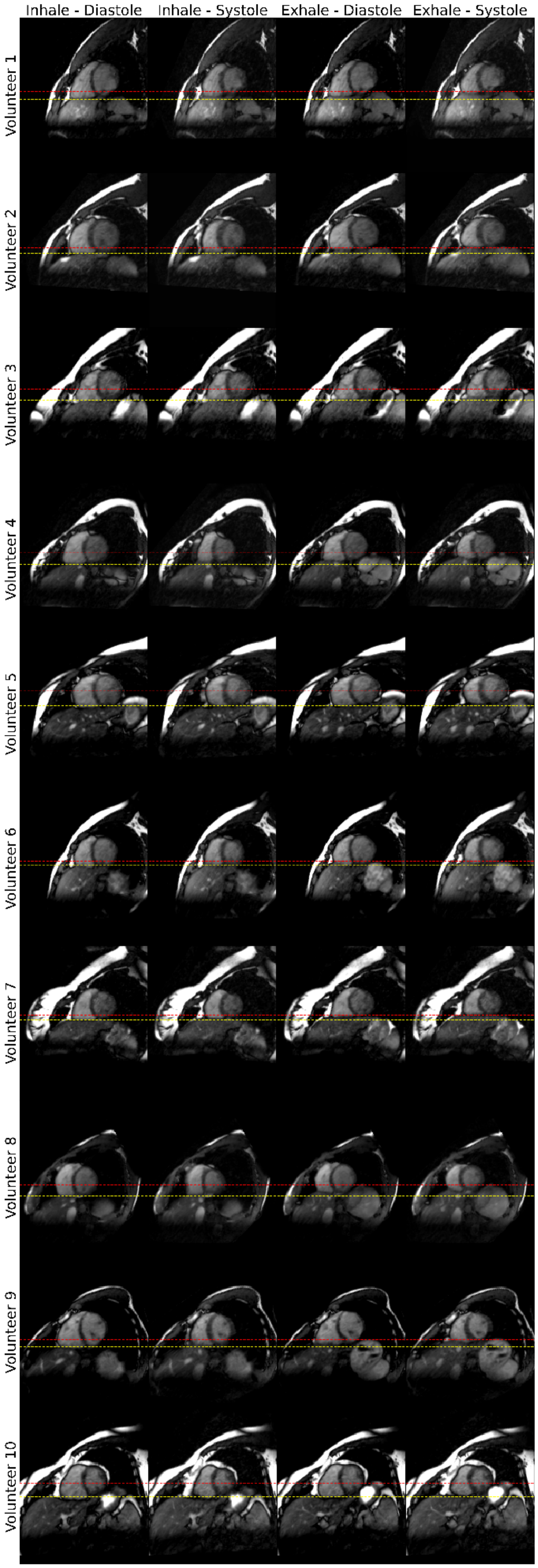

***Figure 6: Reconstruction overview**. Here, an overview of the short-axis reconstructions for all 10 volunteers across four extreme motion states (Diastole/inhale, systole/inhale, diastole/exhale, and systole/exhale) is shown, illustrating motion-corrected images and cardiorespiratory disentanglement between the motion states. An animated figure showing all motion states is provided in Supplementary Figure 1.*

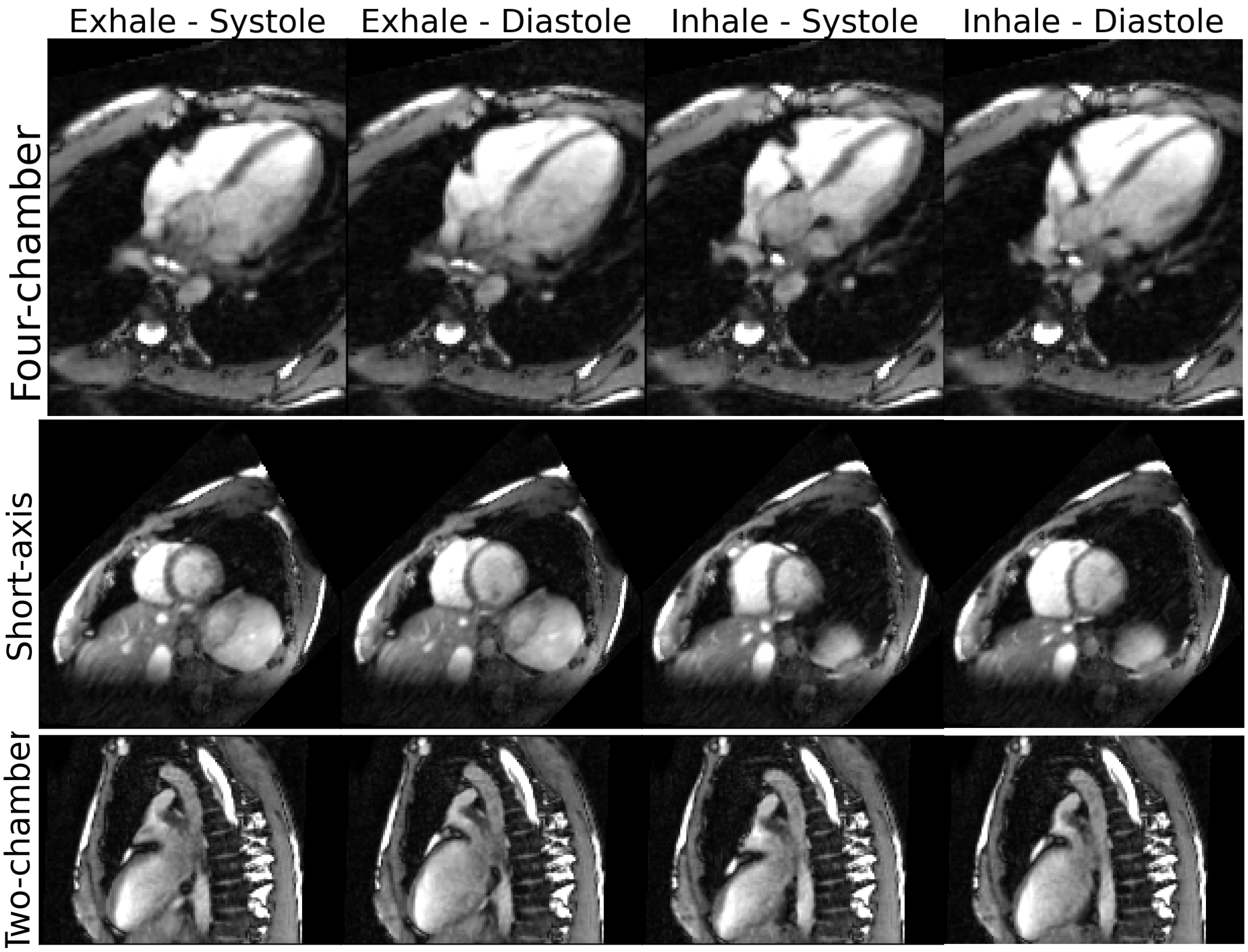

***Figure 7: Cardiac orientations**. The four-chamber, short-axis, and two-chamber orientations are shown for one volunteer across four extreme motion states (Diastole/inhale, systole/inhale, diastole/exhale, and systole/exhale), demonstrating high image quality and motion disentanglement between motion states. An animated figure showing all motion states is provided in Supplementary Figure 2.*

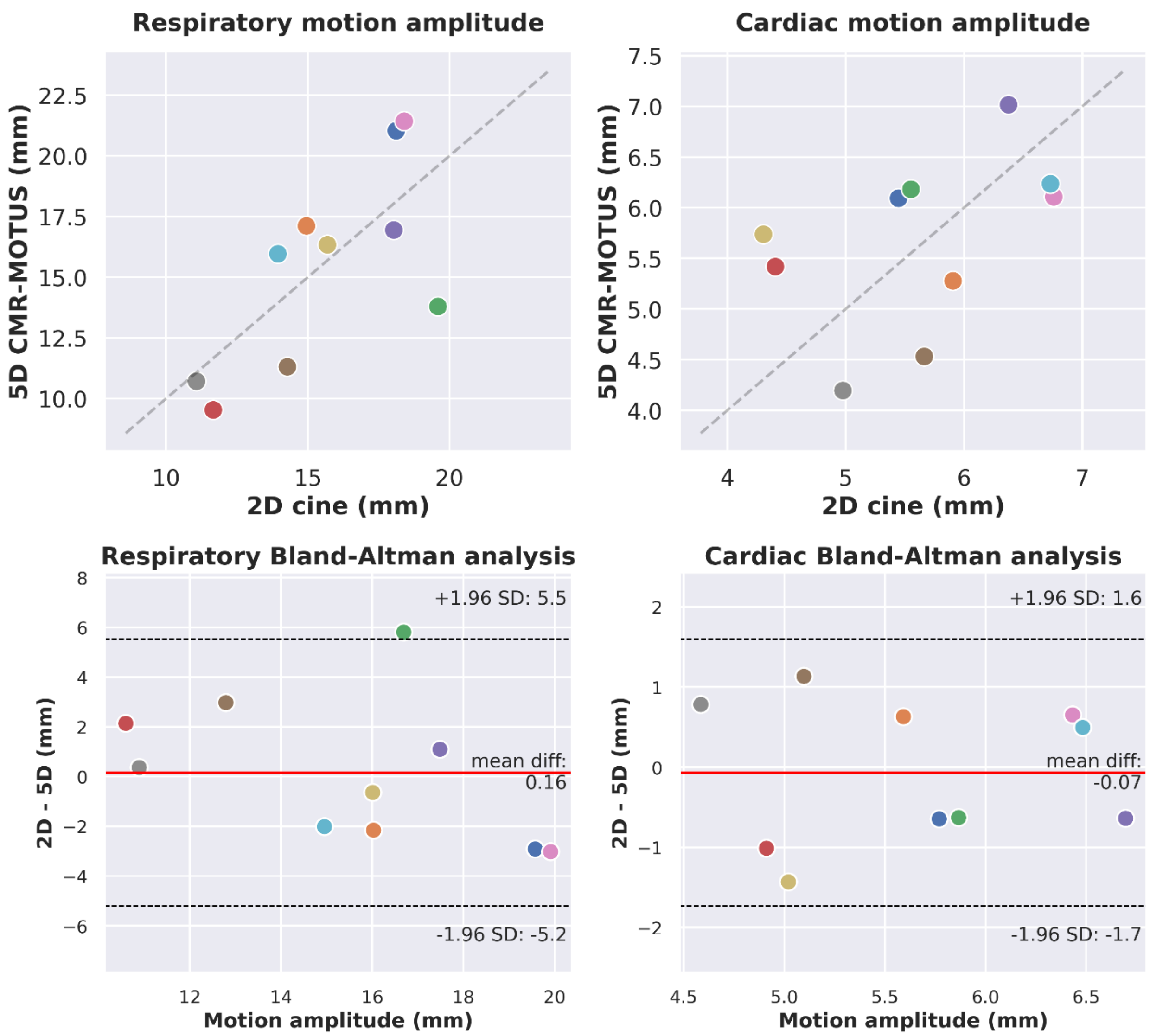

***Figure 8: Quantitative evaluation***. *The top row shows the correlation between the 2D and 5D motion amplitude and the bottom row shows a Bland-Altman analysis. The 5D reconstructions show high agreement with the 2D cine MRI, resulting in a low bias (less than 0.2 mm) and a cardiac standard deviation smaller than the 2 mm voxel size. Identical colors indicate identical volunteers.*

# Discussion

We have proposed 5D CMR-MOTUS, a new method for 5D-MRI reconstruction that captures cardiorespiratory motion without gating or ECG, using a 3D Cartesian acquisition and explicitly modeling low-rank motion fields. We have demonstrated that accurate 5D-MRI can be obtained with a 60-second acquisition on a 1.5T MR-linac and 6 minutes of reconstruction, yielding a total end-to-end latency of approximately 7 minutes for pre-beam motion cardiorespiratory motion characterization. This is significantly faster than other 5D methods, which rely on longer non-Cartesian acquisitions and image-based reconstruction [19–22]. In contrast, our method employs a Cartesian acquisition, obviating the need for computationally expensive non-uniform FFTs, and exploits the intrinsic high compressibility of deformation vector fields compared to time-resolved images, resulting in fast reconstruction. These properties make this method especially suited for online 5D motion characterization, margin estimation, and treatment planning for MRI-guided STAR treatments.

By reconstructing the motion states based on the deformation vector field, several limitations of k-space sorting-based methods are addressed: there is no intra-bin motion due to k-space mismatches, the number of cardiac and respiratory phases can be selected a posteriori instead of an a priori-chosen reconstruction hyperparameter. Moreover, it maximizes the SNR of the reconstruction as all k-space is used to reconstruct every cardiorespiratory motion state. However, the quality is limited by the reference image and the motion fields. Moreover, it assumes that cardiac and respiratory motion are perfectly separable, whereas cardiothoracic motion coupling can prevent this separation [40]. Finally, the framework assumes that all physiological effects can be captured by smooth deformation vector fields, which fails for blood flow or sliding motion [41,42]. Several extensions to the 5D CMR-MOTUS framework can be applied to address these limitations. Future work will investigate these extensions to the 5D CMR-MOTUS framework, such as capturing blood flow effects, evaluating the suitability of bases other than cubic B-splines for non-smooth motion, and using plug-and-play neural networks to further improve the quality of the reference image [43]. Using these pre-trained deep neural networks as denoisers in iterative algorithms has been demonstrated to significantly lower the required convergence time. However, 5D CMR-MOTUS requires GPU acceleration to achieve low reconstruction times, necessitates low-latency, high-throughput streaming solutions to transfer raw k-space to dedicated reconstruction hardware[44].

For this study, we have used a 3D Cartesian acquisition with the OPRA sampling pattern [30]. In general, the 5D CMR-MOTUS framework does not require a particular sampling strategy and could be combined with, for example, a non-Cartesian acquisition. However, such acquisitions are generally more sensitive to system imperfections, such as gradient nonlinearities, and are subject to eddy-current-induced effects [45]. Moreover, the requirement of a non-uniform fast Fourier transform induces a computational burden during iterative optimization [23]. On the other hand, OPRA provided a good trade-off between reduced computational cost, temporal incoherency, and greater resilience to system imperfections due to its minimal jumps in k-space.

We have demonstrated that 5D CMR-MOTUS accurately captures cardiorespiratory motion using digital and physical phantoms. For MRI-guided STAR, motion quality and target visibility are paramount. Currently, we use frequency-based filtering to separate cardiac from respiratory motion. While a 0.7–2.0 Hz cardiac bandpass filter was used for this volunteer cohort, this might be insufficient for patients suffering from ventricular tachycardia, which can cause heart rate to fall outside this window. However, STAR planning and delivery are performed during baseline sinus or paced rhythm rather than sustained rapid VT. Moreover, the bandpass filter is not fixed, and our 10.1 Hz OPRA leaflet sampling rate provides a Nyquist frequency above 5.0 Hz, allowing the upper frequency cutoff to be widened to 3.5–4.0 Hz (210–240 bpm) without overlapping the respiratory band (0.1–0.4 Hz). Furthermore, because the central $k_0$ readout can be sampled in every leaflet as a self-navigation signal, transient arrhythmic episodes or PVCs can be identified from sudden spectral shifts, enabling the exclusion or re-acquisition of affected k-space leaflets from the optimization.

For these acquisitions, we have used a bSSFP pulse sequence. While this pulse sequence provides a low repetition time and high contrast between the blood pool and the myocardium, it is highly sensitive to $B_0$ inhomogeneities, leading to severe susceptibility artifacts in the presence of cardiac implantable electronic devices (CIEDs) [46,47]. Moreover, the high RF power associated with bSSFP acquisitions can lead to SAR concerns for MR-conditional implant devices. However, these changes might be mitigated when considering, for example, gradient-recalled echo sequences, as these also have a low repetition time, reduced SAR due to lower flip angles, and greater robustness to magnetic field inhomogeneity. While exploratory non-contrast 3D GRE OPRA datasets were acquired in our volunteer cohort, the 5D CMR-MOTUS reconstruction settings in this study were tuned exclusively for bSSFP and remain unvalidated for GRE. Without exogenous contrast, spoiled GRE acquisitions yield a lower SNR and lower blood–myocardium contrast than bSSFP, which significantly reduces the cardiorespiratory motion estimation performance. In the future, we will investigate the performance of 5D CMR-MOTUS using GRE acquisitions in a patient population.

In the volunteer study, we have shown high consistency in cardiac and respiratory motion amplitudes between 5D and 2D cine MRI acquisitions. While cardiac motion differences were small ($-0.07 \pm 1.7$ mm), respiratory amplitudes showed a larger standard deviation of 5.5 mm between the 2D and 5D sequences. The duration-matched segment analysis showed that breath-to-breath variability within the 5D scan was small (CV 8%). However, volunteers demonstrated deeper breathing during the 2D acquisition than the 5D acquisition (mean excess 2.9 ± 2.4 mm). This change in respiratory pattern therefore explains part of the increased standard deviation. In general, a fundamental limitation of in-vivo 4D/5D-MRI evaluation is the absence of a ground-truth for in-vivo 3D time-resolved DVFs. Surrogate in-vivo acquisitions (such as 2D cine MRI or 5D-MRI acquisitions reconstructed with alternative methods) are acquired non-simultaneously, in fewer spatial dimensions, or suffer from other reconstruction limitations, such as intra-bin motion or over-regularization along the cardiorespiratory motion dimension.

Finally, we used an imaging resolution of 2 x 2 x 2 $mm^3$ and an acquisition duration of 60 seconds. With these acquisition parameters, accurate 5D-MRI was obtained. While this spatial resolution is sufficient for stereotactic radiation treatments [48], this is a lower resolution than is typically obtained in 5D-MRI for radiological examinations [19,21,22]. However, the large field of view requirement for MRI-guided STAR treatments to visualize the body contour and the latency requirements reduce the maximum achievable resolution. On the other hand, the pulse sequence parameters and the trade-off between acquisition time, reconstruction time, and image quality were not exhaustively optimized for in vivo acquisitions. Future studies will investigate the optimal sampling strategy to maximize motion accuracy and target visibility.

# Conclusion

We have presented 5D CMR-MOTUS, a novel 5D-MRI reconstruction method that combines a 3D Cartesian acquisition low-rank deformation vector fields that explicitly disentangle the respiratory and cardiac motion. We have shown that it is feasible to obtain accurate 5D-MRI with a total latency of seven minutes (one minute acquisition and six-minute reconstruction). This is particularly interesting for MRI-guided STAR treatments, where 5D-MRI must be available as quickly as possible. The method was extensively validated using phantoms and demonstrated accurate 5D-MRI reconstructions. The reconstructions and k-space are publicly available at https://doi.org/10.5281/zenodo.21278894.

# Acknowledgments

We gratefully acknowledge the help of Katrinus Keijnemans and Edwin Versteeg for their technical support. This work was supported by the Dutch Research Council (NWO) through the PAMPER (19003) and MEGAHERTZ (19484) projects.

## Author contributions

CRediT: ML Terpstra: Conceptualization, Data curation, Formal Analysis, Investigation, Methodology, Resources, Software, Supervision, Validation, Visualization, Writing – original draft, Writing – review & editing; TE Olausson: Conceptualization, Data curation, Formal Analysis, Investigation, Methodology, Resources, Software, Validation, Visualization, Writing – original draft, Writing – review & editing; M Aubert: Data curation, Investigation, Project administration, Resources, Software, Validation, Writing – review & editing; C Beijst: Funding acquisition, Methodology, Project administration, Resources, Supervision, Writing – review & editing; A Sbrizzi: Conceptualization, Funding acquisition, Methodology, Project administration, Resources, Supervision, Writing – review & editing; CAT van den Berg: Conceptualization, Funding acquisition, Methodology, Project administration, Resources, Supervision, Writing – review & editing; MF Fast: Conceptualization, Funding acquisition, Methodology, Project administration, Resources, Supervision, Writing – review & editing

## Financial disclosure

## Conflict of interest

The authors declare no potential conflict of interests.

# Supporting information

The following supporting information is available as part of the online article:

**Supplemental Figure 1. Animated version of Figure 6.**

**Supplemental Figure 2. Animated version of Figure 7.**

Both supplementary figures and Supplementary Information Document 1 can be found at https://doi.org/10.5281/zenodo.21293931.